\documentclass[twoside,a4paper,11pt]{sca}
\usepackage{graphicx}
\usepackage{hyperref}
\usepackage{natbib}  % Cross-reference package (Natural BiB)
\begin{document}
\pagenumbering{arabic}
\pagestyle{myheadings}
\thispagestyle{empty}
%%%%{\flushright\includegraphics[width=\textwidth,bb=58 650 590 680]{stamp.pdf}}
{\flushright\includegraphics[width=\textwidth,bb=90 650 520 700]{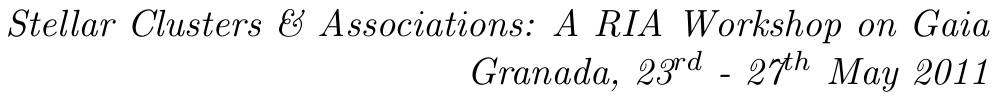}}
\vspace*{0.2cm}
\begin{flushleft}
{\bf {\LARGE
%
%%% TITLE of the paper. 
%%% TITLE of the paper. 
Dissecting high-mass star-forming regions; tracing back their complex formation history
%
% Do not delete next few lines
}\\
\vspace*{1cm}
%
%%% Include here the LIST OF AUTHORS.
%%% Include here the LIST OF AUTHORS.
%%% Note that the last author has to be preceeded by an AND.
Arjan Bik$^{1}$,
Thomas Henning$^{1}$, 
Andrea Stolte$^{2}$,
 Wolfgang Brandner$^{1}$,
 Dimitrios Gouliermis$^{1}$,
Mario Gennaro$^{1}$,
Anna Pasquali$^{3}$,
Boyke Rochau$^{1}$,
Henrik Beuther$^{1}$,
and
Yuan Wang$^{4}$
%
% Do not delete next few lines
}\\
\vspace*{0.5cm}
%
%%% AFFILIATIONS LIST.
%%% and the AFFILIATIONS LIST. Note that one affiliation per line.
%%% Add as many affiliations as necessary. 
$^{1}$
Max-Planck-Institut f\"ur Astronomie, K\"onigstuhl 17, 69117 Heidelberg, Germany\\
$^{2}$
Argelander Institut f\"ur Astronomie, Auf dem H\"ugel 71, 53121 Bonn, Germany\\
$^{3}$
Astronomisches Rechen Institut, M\"onchhofstrasse 12 - 14, 69120 Heidelberg, Germany\\
$^{4}$
Purple Mountain Observatory, Chinese Academy of Sciences, 210008, Nanjing, PR China
%
% Do not delete next few lines
\end{flushleft}
%
% Headings
\markboth{
%%% Type the SHORT version of the paper title.
%%% Type the SHORT version of the paper title.
Dissecting high-mass star-forming regions
}{ % Do not delete
%
%%%  First Author \& Second Author   OR   First-author et al. 
%%%  First Author \& Second Author   OR   First-author et al. if the author list 
%%% contains three or more authors.
Bik et al. 
% 
% Do not delete next few lines
}
\thispagestyle{empty}
\vspace*{0.4cm}
\begin{minipage}[l]{0.09\textwidth}
\ 
\end{minipage}
\begin{minipage}[r]{0.9\textwidth}
\vspace{1cm}
\section*{Abstract}{\small
%
% ABSTRACT ABSTRACT ABSTRACT
% ABSTRACT ABSTRACT ABSTRACT
%%% Type the ABSTRACT of your paper
We present near-infrared JHKs imaging as well as K-band multi-object spectroscopy of the massive stellar content of W3 Main using LUCI at the LBT. We  confirm 13  OB stars by their absorption line spectra in W3 Main and spectral types between O5V and B4V have been found. Three massive Young Stellar Objects are  identified by their emission line spectra and near-infrared excess.  
From our spectrophotometric analysis of the massive stars and the nature of their surrounding HII regions we derive the evolutionary sequence of W3 Main and we find evidence of an age spread of at least 2-3 Myr. While the most massive star (IRS2) is already evolved, indications for high-mass pre--main-sequence evolution is found for another star (IRS N1), deeply embedded in an ultra compact HII region, in line with the different evolutionary phases observed in the corresponding HII regions. We have detected the photospheres of OB stars from the more evolved diffuse HII region to the much younger UCHII regions, suggesting that the OB stars have finished their formation and cleared away their possible circumstellar disks very fast. Only in the hyper-compact HII region (IRS5), the early type stars are still surrounded by circumstellar material.
%
% Do not delete next few lines
\normalsize}
\end{minipage}
%
%
%%% BODY of the paper
%%% BODY of the paper
%
\section{Introduction \label{intro}}

Despite the impact on their surroundings, the formation and early evolution of massive stars is poorly constrained, primarily because of their scarcity and short lifetimes.  Most OB stars form in stellar clusters and associations, and are still partly hidden in their natal molecular cloud.  To detect the stellar content of these regions directly, near-infrared observations, especially spectroscopy, have proven to be a powerful method to find and characterize the newborn OB stars.  A pure photometric characterization of young stellar clusters is strongly hampered by highly varying extinction, unknown distances and infrared excess of the young cluster members.

A spectroscopic identification, however,  results in a unambiguous identification of the (massive) stellar content. Stellar properties, like effective temperature and luminosity, are derived based on the spectral features and extinction, distance and infrared excess can be determined reliably \citep[e.g.][]{Blum00, Hanson02,Ostarspec05,Puga06,Bik10,Puga10}. By comparing the effective temperature and  luminosity to main sequence and pre-main-sequence isochrones, the age of a stellar cluster can be derived more reliably  than from photometry alone \citep{Bik10}.  When studying larger samples of massive stars in the same star cluster, estimates  of an age spread in the cluster can be obtained.  Based on the massive stars, \citet{Clark05} finds very little age spread (less than 1 Myr) in starburst clusters. However, in Orion an age spread has been found based on the PMS stars \citep{daRioOrion10}. Additional evidence for an age spread has been found in S255 where the most massive star is still deeply embedded and driving an outflow, while the PMS population has an age of 1-3 Myr \citep{Wang11}. 

We present deep LUCI near-infrared JHKs imaging as well as K-band multi-object-spectroscopy of the massive stellar content of W3 Main. This allows for the first time a spectral classification of the massive stellar content. Using their derived spectral type we discuss the evolutionary status of the massive stars as well as the HII regions in  detail. The different evolutionary phases of the HII regions make W3 Main an ideal target to study age spread and the evolution of circumstellar disks around massive stars.

\section{Near-infrared imaging and spectroscopy}

\begin{figure}
\center
\includegraphics[width=0.465\textwidth]{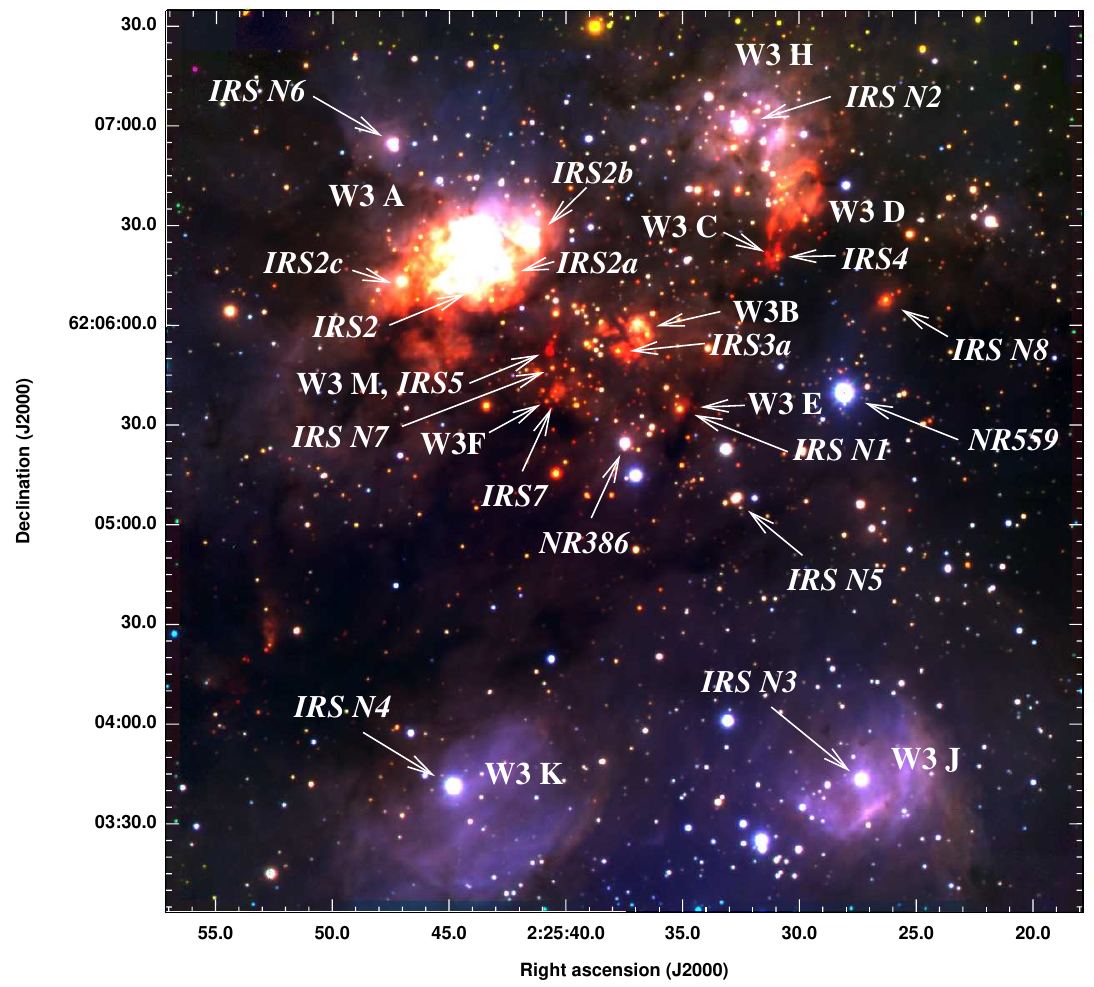}~
\includegraphics[width=0.535\textwidth]{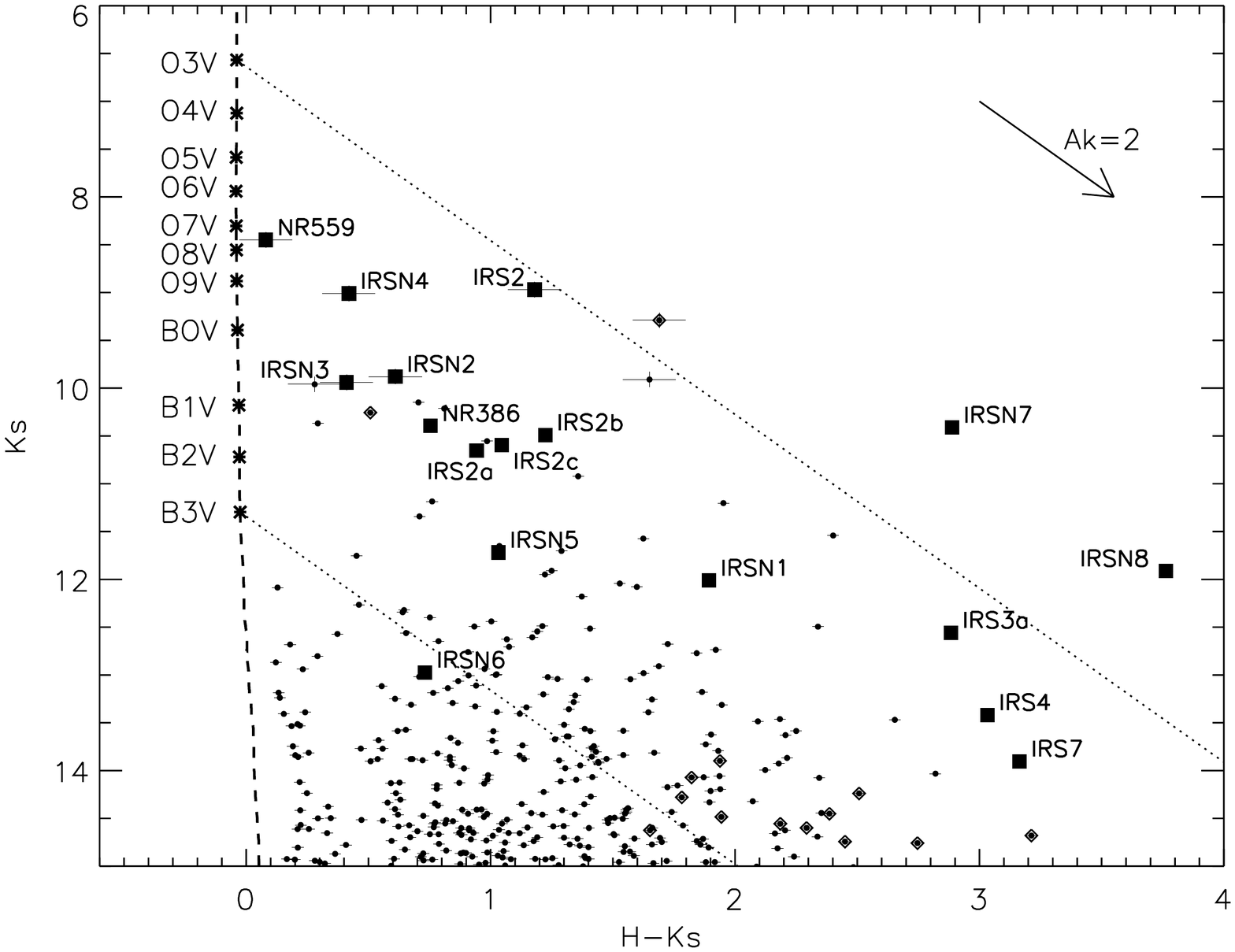}
\caption{\label{fig:image} \emph{Left:} JHKs color composite of W3 Main taken with LUCI at the LBT. The HII regions as well as the massive stars are annotated. \emph{Right:} Overplotted as a dashed line is the main sequence isochrone with an age of 1 Myr from \citet{LeJeune01} as well as the location of the spectral types between O3V and B3V are indicated. 
The two diagonal dotted lines show the reddening lines for an O3V and a B3V star using the extinction law of \citet{Indebetouw05}.  Additionally, the reddening vector for A$_{Ks}$=2 mag is indicated.}

\end{figure}

JHKs imaging of W3 Main has been obtained with LUCI at the Large Binocular Telescope (LBT) on Mount Graham, Arizona. Fig. \ref{fig:image} shows the resulting three color image of a 4' $\times$ 4' field centered on W3 Main. Photometry on the JHKs images was performed using the \emph{starfinder} software  \citep{Diolaiti00}. The photometry as well as the astrometry is calibrated with 2MASS. The resulting Ks vs H-Ks color-magnitude diagram (Fig. \ref{fig:image} ) shows the enormous spread in color of the massive stars caused by large extinction variations. 

K-band spectra of  16 candidate massive stars inside W3 Main were obtained with the multi-object spectroscopic mode of LUCI. We classified 13 stars as OB stars from their absorption lines (Fig. \ref{fig:hrd}) compared with  K-band spectra of optically visible OB stars \citep{Hanson05,Ostarspec05}  Three objects are classified as massive Young Stellar Objects, massive stars surrounded by a circumstellar disk. See Bik et al. (to be submitted) for a detailed description of the obtained data and reduction process.

\section{HR-diagram}

\begin{figure}
\center
\includegraphics[width=0.535\textwidth]{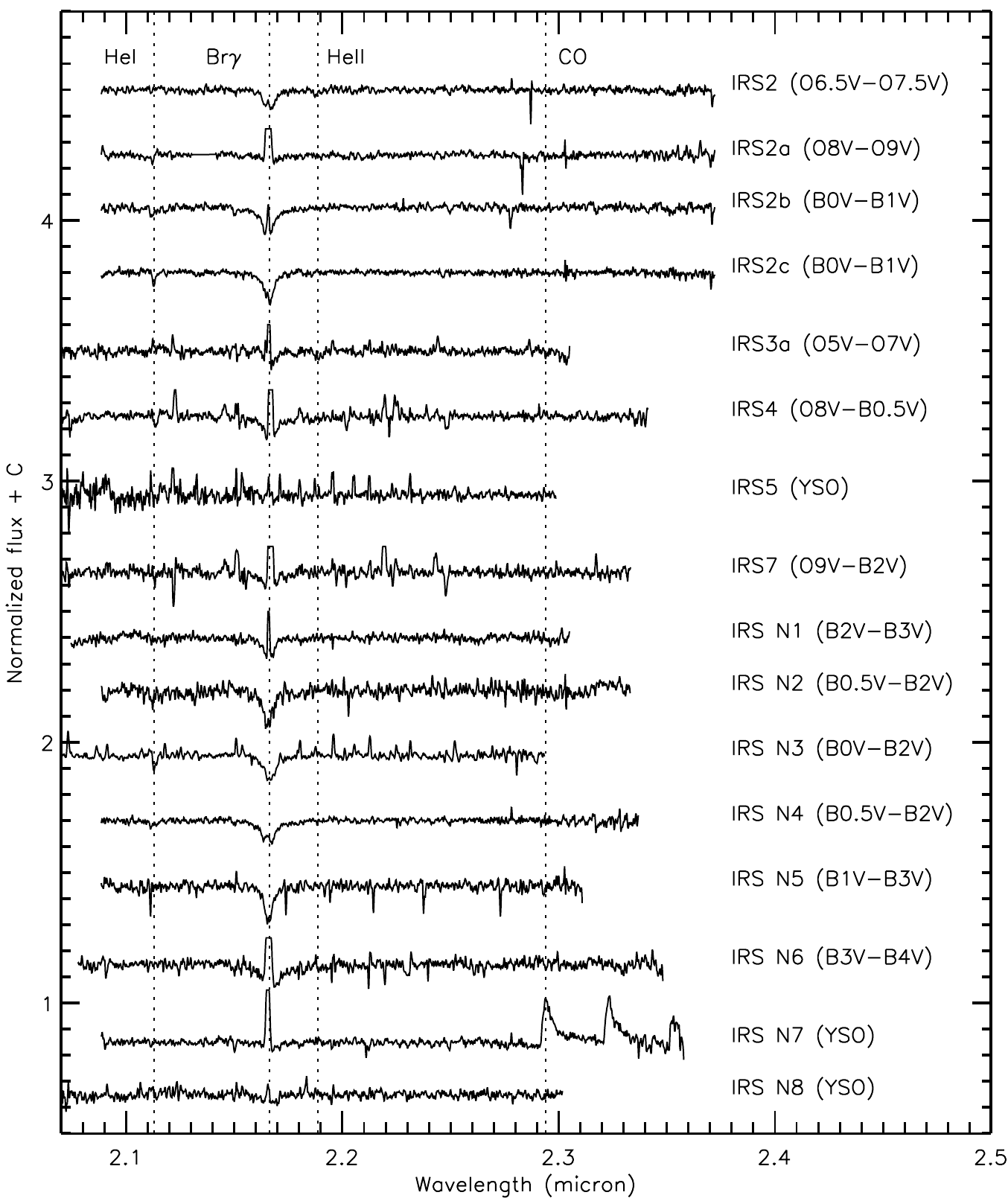}~
\includegraphics[width=0.465\textwidth]{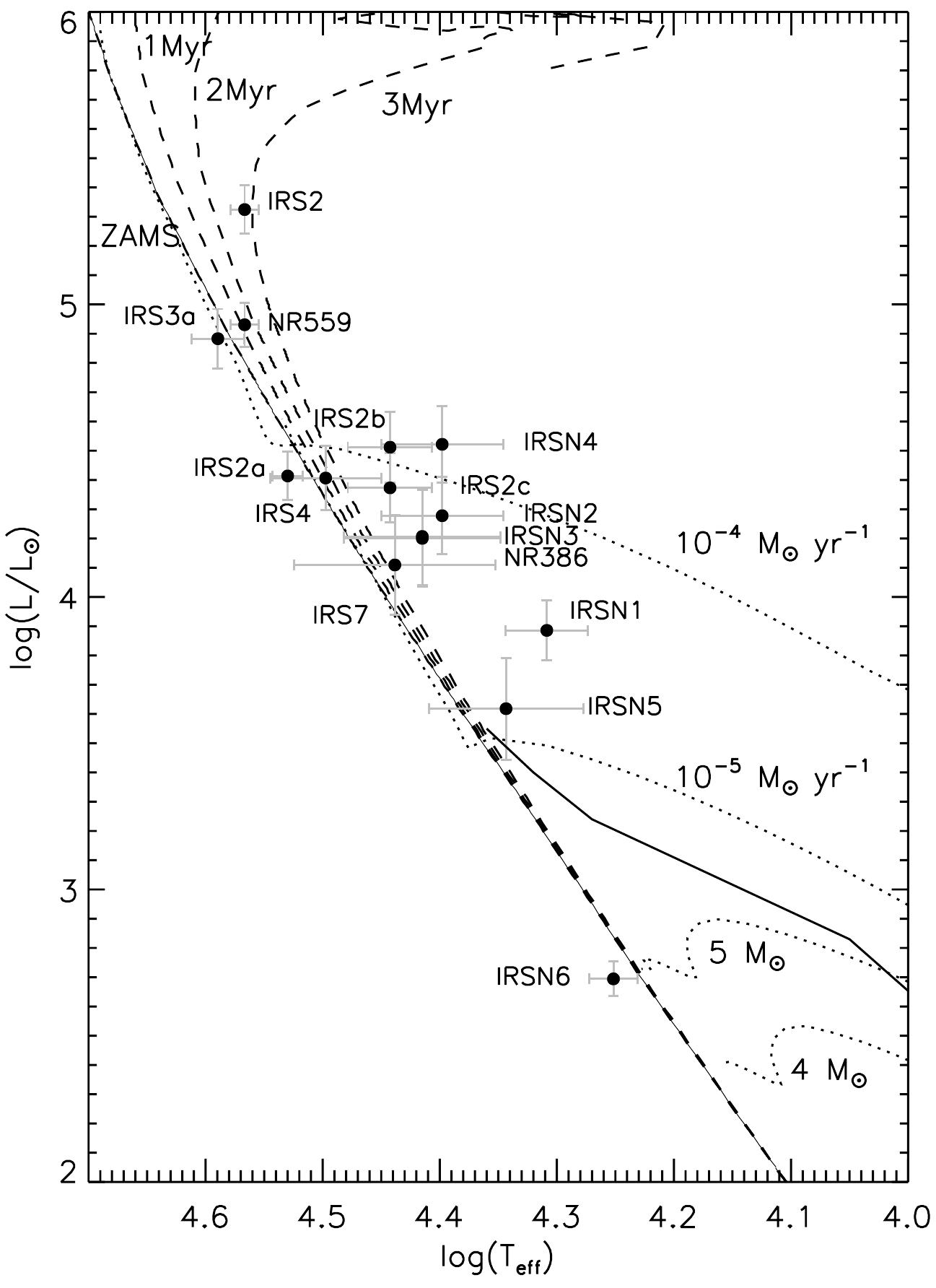}
\caption{\label{fig:hrd} \emph{Left:}  Normalized K-band spectra of the massive stars in W3 Main as taken with the multi-object-mode of LUCI. Annotated with dashed lines are the spectral features which are important for the classification of their stellar spectra.
\emph{Right:} Hertzsprung Russell diagram  of the massive stars  in W3 Main. The dashed lines represent the main-sequence isochrones from \citet{LeJeune01} for 1, 2 and 3 Myr. The dotted lines show the pre-main-sequence evolutionary tracks for a 4 and 5 M$_{\odot}$ star \citep{Siess00} as well as high-mass protostellar evolution tracks for 10$^{-4}$ and 10$^{-5}$ M$_{\odot}$ yr$^{-1}$ from \citet{Hosokawa09}.  The solid line represents the theoretical birth-line, above which  an optically visible pre-main-sequence star is not detectable \citep{Palla90}. }
\end{figure}

After their spectral classification, the bolometric luminosities of the massive stars are calculated and the stars are placed in the Hertzsprung Russell diagram (HRD, Fig. \ref{fig:hrd}) and compared with stellar evolution isochrones. Overplotted in the HRD are the main sequence isochrones from the zero age main sequence (ZAMS) to 3 Myr from \citet{LeJeune01}, the PMS isochrones for intermediate mass stars  \citep[M $<$ 5 M$_{\odot}$,][]{Siess00} as well as the theoretical birth-line \citep{Palla90}. Overplotted, also, are the high-mass protostellar evolutionary tracks from \citet{Hosokawa09} for 10$^{-5}$ and 10$^{-4}$ M$_{\odot}$ yr$^{-1}$ accretion rates.

In the upper regions of the HRD (log L/L$_{\odot}$ $\geq$ 5), the main sequence isochrones indicate that the most massive stars already show significant evolution  after a few Myr. Two stars are located in the upper regions of the HRD, IRS2 and IRS3a.  IRS2 is located to the right of the ZAMS and its location is more consistent with the 2-3 Myr isochrones. The foreground extinction is not very extreme, therefore the  location of IRS2 in the HRD does not vary a lot by changing the extinction law. IRS3a (O5V -- O7V), however, is very reddened (A$_{Ks}$=5.42 $\pm$ 0.79 mag) and its location in the HRD depends very strongly on the adopted extinction law. Additionally, the presence of a near infrared excess makes its luminosity more uncertain. However, the effective temperature determination is more reliable and independent of extinction and excess. This allows us to place an upper limit on the age of 3 Myr, as stars on older isochrones will be of later spectral type than IRS3a.

In the middle area of the  HRD (3.5 $\leq$ log L/L$_{\odot}$ $\leq$ 5), most of the sources are within 3$\sigma$ of the main sequence. For spectral types between O9V and B1V, where most of the objects are found, the change in effective temperature as well as K-band magnitude is very large, while the change in spectral features is not that dramatic.  This is reflected in the large error bars of e.g. IRS N5 and IRS7.  However IRS N1 and also IRS N4 are significantly away from the main sequence. As their extinction is relatively low, their position does not change much with extinction. Below we discuss possible scenarios explaining this offset.

Most of the massive stars are observed to be binary stars \citep[e.g.][]{Apai07}. An equal mass binary  would result in a mismatch in the observed luminosity of a factor of 2 (log L/L$\_{\odot}$  = 0.3), with a multiple star system, increasing the offset even more. Both sources would have to be a system of 5-6 equal mass stars to explain their location above the main sequence. 

Stellar evolution in main sequence or pre-main-sequence could be the reason why these objects are located away from the main sequence. The location of IRS N4 would be compatible with a 10 Myr main sequence track, while IRS N1 would require an even older isochrone. This is highly unlikely for IRS N1, as the star is located inside an ultra-compact HII region,  still very young \citep[$10^5$ years,][]{WoodIRAS89}. IRS N4 is the central star of a diffuse and more evolved HII region, however, as shown by \citet{Lada03}, a time span of 10 Myrs would be enough to disperse the surrounding gas.

Above the "birth-line" in the HRD, no optically visible PMS stars are expected \citep{Palla90}.  However, using deep near-infrared observations we might start to probe the pre-main-sequence phase of the massive stars. Theoretical modeling of  high-mass proto-stellar evolution predicts that massive proto-stars might have a similar evolution to intermediate PMS stars towards the main sequence \citep{Hosokawa09}. The exact path of the proto-stars in the HRD depends on their accretion rate. Overplotted in Fig. \ref{fig:hrd} are two Hosokawa tracks for  10$^{-4}$ M$_{\odot}$ yr$^{-1}$ and 10$^{-5}$ M$_{\odot}$ yr$^{-1}$. The location of IRS N1 would be consistent with  an accretion rate in between 10$^{-5}$ and 10$^{-4}$ M$_{\odot}$ yr$^{-1}$.

\section{Discussion and Conclusions}

W3 Main harbors several different evolutionary stages of HII regions, ranging from very young hyper-compact HII (HCHII) regions (few 10$^3$ years), ultra-compact HII (UCHII) regions \citep[$\sim 10^5$ years][]{WoodIRAS89} to evolved,  diffuse HII regions (few 10$^6$ years). All these regions are most likely formed out of the  same molecular cloud. This provides the possibility to study the evolution of young HII regions and their stellar content in great detail.  \citet{Tieftrunk97} derived an evolutionary sequence for the HII regions in W3 Main, based on the morphology of the radio sources.  
The youngest are the HCHII regions W3 M and W3 Ca, with the UCHII regions W3 F, W3 C and W3 E, slightly older, the compact HII regions W3 B and W3 A even more evolved, and the diffuse HII regions, W3 K and W3 J being the oldest HII regions in W3 Main. This classification can be compared to the ages of the massive stars deduced from their position in the HRD. 

We have detected OB stars in three diffuse HII regions, two compact HII regions and three UCHII regions. The HCHII region W3 M harbors the high-mass protostar IRS5 (see Sect. 4.1).  Additionally three stars have no detectable HII region associated with them (IRS N5 and the massive YSOs IRS N7 and IRS N8).  The position of IRS2 in the HRD suggests an age of  2-3 Myr, consistent with its location in a relatively evolved compact HII region (W3 A) with a similar or younger age derived for  IRS3a,  harbored in a younger UCHII region. For the lower-mass stars (late O, early B) the isochrones are too close to each other to derive any age information. For IRS N1, located inside the UCHII region W3 E, the offset from the main sequence could be explained by high-mass pre-main-sequence evolution \citep{Hosokawa09}, consistent with the expected young age of the UCHII region.
 
A similar sequence can be seen in the extinction towards the different HII regions. The extinction varies from A$_{Ks}$ = 0.9 mag for the diffuse HII regions W3 J and W3 K to very high extinction (A$_{Ks}$  = 5.9 mag) towards  W3 F.  Such a sequence in extinction indicates that the stars in the diffuse HII regions have already cleared out their surroundings and destroyed the molecular cloud, while those in UCHII regions are still  embedded in their parental molecular cloud.

The evolutionary sequence of the HII regions seen in the radio morphology, is consistent with an increase in extinction from the older (diffuse HII) regions to the younger (UCHII) regions.  We have detected the photospheres of OB stars from the more evolved diffuse HII region to the much younger UCHII regions, suggesting that the OB stars have finished their formation and cleared away their possible circumstellar disks very fast. Only in the HCHII phase (IRS 5), the massive star is still surrounded by circumstellar material.

Based on the presence of different evolutionary stages of HII regions as well as the location of the most massive stars in the HRD, we can conclude that an age spread of  2-3 Myr is most likely present for the massive stars in W3 Main. A growing number of  young stellar clusters show evidence for an age spread, usually based on the analysis of their PMS population in the HRD. In Orion an age spread of a few Myr has been found  \citep{Palla99,daRioOrion10}. In starburst clusters, however, upper limits on the age spread of less than 1 Myr have been found for Westerlund 1 \citep{Clark05} based on its massive stars. This suggests that W3 Main is not formed in one star formation burst as expected for starburst cluster, but, more likely, through a temporal sequence of star formation events.

%
%
% Do not delete the next line
\small  % Do not delete
%
%%% Comment the following line if you do not have acknowledgments.
%\section*{Acknowledgments}   % Do not delete if you declare acknowledgments
%%
%%%% ACKNOWLEDGMENTS
%%%% ACKNOWLEDGMENTS
%If you do not have any acknowledgments, you may comment this Section.

%
% Do not delete the next few lines
%************************************************************************************%

%************************************************************************************%
%
% Do not delete the next few lines

%\bibliographystyle{aa}
%\bibliography{mnemonic,Bik_A.bib}

\end{document}